\documentclass[twocolumn]{aa}
\usepackage{graphicx}
\usepackage{natbib,graphics}
\usepackage{times}
\bibpunct{(}{)}{;}{a}{}{,}
\begin{document}
\title{\object{2MASS\,J0516288+260738}: Discovery of the first \\
    eclipsing late K + Brown dwarf
    binary system?\thanks{This paper uses observations made at the Bohyunsan Optical Astronomy
    Observatory of Korea Astronomy Observatory, at the South African
    Astronomical Observatory (SAAO), at the 0.9\,m telescope at Kitt Peak
    National Observatory recommissioned by the Southeastern Association for
    Research in Astronomy (SARA), at Gunma Astronomical Observatory
    established by Gunma prefecture, Japan, at the Florence and George
    Wise Observatory, operated by the Tel-Aviv University, Israel and
    at Piszk\'estet\H o, the mountain station of Konkoly Observatory of the
    Hungarian Academy of Science, Hungary.}\fnmsep\thanks{This
    publication makes use of data products from the Two Micron All Sky Survey,
    a joint project of the University of Massachusetts and the Infrared
    Processing and Analysis Center / California Institute of Technology,
    funded by the National Aeronautics and Space Administration and the
    National Science Foundation.}\fnmsep\thanks{The Digitized Sky
    Survey was produced at the Space Telescope Science Institute under
    U.S. Government grant NAG W-2166. The images of these surveys are
    based on photographic data obtained using the Oschin Schmidt
    Telescope on Palomar Mountain and the UK Schmidt Telescope. The
    plates were processed into the present compressed digital form
    with the permission of these institutions.
}}
\author{S.L.~Schuh\inst{1,14}\fnmsep\thanks{Visiting Astronomer, German-Spanish
    Astronomical Centre, Calar Alto, operated by the
    Max-Planck-Institute for Astronomy, Heidelberg, jointly with the
    Spanish National Commission for Astronomy.}
  \and
  G.~Handler\inst{2,3}
  \and
  H.~Drechsel\inst{4}
  \and
  P.~Hauschildt\inst{5}
  \and
  S.~Dreizler\inst{1,14}\fnmsep$^\dagger$
  \and
  R.~Medupe\inst{3,6}
  \and
  C.~Karl\inst{4}\fnmsep$^\dagger$
  \and
  R.~Napiwotzki\inst{4}\fnmsep$^\dagger$
  \and
  S.-L.~Kim\inst{7}
  \and
  B.-G.~Park\inst{7}
  \and
  M.A.~Wood\inst{8}
  \and
  M.~Papar\'o\inst{9}
  \and
  B.~Szeidl\inst{9}
  \and
  G.~Vir\'aghalmy\inst{9}
  \and
  D.~Zsuffa\inst{9}
  \and
  O.~Hashimoto\inst{10}
  \and
  K.~Kinugasa\inst{10}
  \and
  H.~Taguchi\inst{10}
  \and
  E.~Kambe\inst{11}
  \and
  E.~Leibowitz\inst{12}
  \and
  P.~Ibbetson\inst{12}
  \and
  Y.~Lipkin\inst{12}
  \and
  T.~Nagel\inst{1}\fnmsep$^\dagger$
  \and
  E.~G\"ohler\inst{1}\fnmsep$^\dagger$
  \and
  M.L.~Pretorius\inst{13}
}
\offprints{S.L.~Schuh}
\institute{Institut f\"ur Astronomie und Astrophysik, Universit\"at
  T\"ubingen, Sand 1,  D--72076, T\"ubingen, Germany,\\
  \email{schuh@astro.uni-tuebingen.de}
  \and
  Institut f\"ur Astronomie, Universit\"at Wien, T\"urkenschanzstra\ss e
  17, A--1180 Wien, Austria
  \and
  South African Astronomical Observatory, PO Box 9, Observatory 7935, Cape, South Africa
  \and
  Dr.-Remeis-Sternwarte, Astronomisches Institut der
  Universit\"at Erlangen-N\"urnberg, Sternwartstr.~7, D--96049 Bamberg, Germany
  \and 
  Hamburger Sternwarte, Universit\"at Hamburg, Gojenbergsweg 112,
  D--21029 Hamburg, Germany
  \and
  Department of Physics, University of the North-West,
    Private Bag X2046, Mmabatho 2735, South Africa
 \and
  Korea Astronomy Observatory, 61-1, Whaam, Yuseong, Daejeon, 305-348,
  Korea
  \and
  Department of Physics and Space Sciences and SARA Observatory, 
  Florida Institute of Technology, 150 West University Boulevard, Melbourne, 
  FL~32901-6975, USA
  \and
  Konkoly Observatory, Box 67, H-1525 Budapest XII, Hungary
  \and
  Gunma Astronomical Observatory, 6860-86 Nakayama Takayama-mura
  Agatsuma-gun Gunma-ken, Postal Code: 377-0702, Japan
  \and
  Department of Earth and Ocean Sciences, National Defense Academy,
  Yokosuka, Kanagawa 239-8686, Japan
  \and
  Wise Observatory, Sackler Faculty of Exact Sciences, Tel Aviv
  University, Tel Aviv 69978, Israel
  \and
  Department of Astronomy, University of Cape Town, Rondebosch~7700, South
  Africa
  \and
  Universit\"atssternwarte~G\"ottingen, Geismar~Landstra\ss e~11,
  D--37083~G\"ottingen, Germany
}
\date{\vspace{-5mm}}
%
%
\abstract{We report the discovery of a new eclipsing system less than
  one arcminute south of the pulsating DB white dwarf \object{KUV\,05134+2605}. The
  object could be identified with the point source \object{2MASS\,J0516288+260738}
  published by the Two Micron All Sky Survey. We present and discuss the
  first light curves as well as some additional colour and spectral
  information. The eclipse period of the system is 1.29\,d, and,
  assuming this to be identical to the orbital period, the
  best light curve solution yields a mass ratio of m$_2$/m$_1=0.11$, a
  radius ratio of r$_2$/r$_1\approx$1 and an inclination of
  74$^\circ$. The spectral anaylsis results in a T$_{\rm eff}=4\,200$\,K
  for the  primary. On this basis, we suggest that the new system  
  probably consists of a late K + Brown dwarf (which would imply a
  system considerably younger than $\approx$0.01\,Gyr to have
  r$_2$/r$_1\approx$1), and outline possible future observations.
  }
\maketitle
%
\section{Introduction}
Detached eclipsing binaries provide precise fundamental stellar
parameters like mass and radius and are thus the prerequisite for the
validation of stellar evolutionary models. The empirical constraints from over
four dozen systems have shown that for main sequence stars between 1
and 10\,M$_{\sun}$ the agreement is acceptable, i.e. better than 2\%
\citep{1991A&ARv...3...91A,1998IAUS..189...99A}, while at the lower main
sequence the situation is far less satisfying. Up to now, only
three eclipsing systems with M-type primaries are known, despite the fact
that low mass main sequence stars dominate the stellar population by
number. The first such system to be discovered was \object{YY\,Gem}
\citep{1926ApJ....64..250J,1926BAN.....3..121V}, followed by
\object{CM\,Dra} (\citealt{1967ApJ...148..911E}, and references
therein) and \object{CU Cnc} \citep{delfosse:99b}; first mass
determinations came from \citet{leung:78}, \citet{lacy:77} and
\citet{delfosse:99b}, respectively.  While \cite{metcalfe:96} find the
slope of the mass-radius relation derived from the M dwarf binary
system \object{CM\,Dra} in agreement with model predictions,
\citet{2000A&A...364..217D} reported on a disagreement between
  empirical and theoretical mass-luminosity relations of 10-20\% in
  the V band, and recent precise
analyses of \object{YY\,Gem} \citep{2002ApJ...567.1140T} and \object{CU\,Cnc}
\citep{ribas:03} also revealed an underestimation (10-20\%) of the
radii of low mass stars from current evolutionary models. Additional
constraints for the empirical mass-radius relation are provided by the
first interferometric measurements of radii from lower main
sequence stars \citep{segransan:03}. These results agree well with
model predictions at the present accuracy level, with a possible
discrepancy for stars with 0.5-0.8\,M$_{\sun}$. Such observations do not provide an
independent measurement of the stellar mass, however, so that
eclipsing systems still are the primary source for a model-independent
determination of fundamental parameters.
\par
Future improvements of the theoretical mass-radius relation for the
lower main sequence would strongly benefit from a larger empirical
database through an increased sample of eclipsing binaries.  Recently,
137 eclipsing low-luminosity candidates were announced by the OGLE
(Optical Gravitational Lensing Experiment) consortium
\citep{2002AcA....52....1U,2002AcA....52..115U,2003AcA....53..133U},
of which several of the
secondaries turned out to be M-type stars
\citep{2002A&A...391L..17D}.  In this paper we report the discovery
of another interesting eclipsing binary system,
\object{2MASS\,J0516288+260738}, whose components appear to
bracket the M-star range, with the potential of extending the
empirical mass-radius relation into the sub-stellar range.
\par
The new eclipsing system has been discovered in observational data
taken during a coordinated photometric monitoring campaign in December
2001. This dataset has been obtained to monitor the light variations
of the DB variable white dwarf \object{KUV\,05134+2605}
({\citealt{grauer:89}, Handler {et al.} in prep.). It consists of many
individual light curves taken by either photomultiplier (PMT) or CCD instruments; the
newly discovered object is included in 48 individual time series of
images obtained with CCD cameras.  While analysing field stars for
photometric stability to check whether they could be used as
references, an object located a little less than one arcminute south
of the DB was found to show the signature of an eclipse in the Calar
Alto 2.2\,m data set of 2001 Dec 07 (see
Table~\ref{tab:observations}). Subsequent searches in the other data
sets 
revealed that eight more eclipses had partly or fully been observed. A
year later, 5 additional data sets were obtained, two of which covered
the eclipse. The full time-resolved photometric data are compiled in
Section \ref{sec:lightcurve}. Archive searches contributed an
identification of the object as well as additional colour information
(Sections \ref{sec:positional} and \ref{sec:colour+spectrum}). Two
months after the initial observations, an optical spectrum could be
obtained, and in the following observing season, an infrared spectrum
was taken (see Section \ref{sec:colour+spectrum}).
\par
In the following, we compile the information that is currently available on
the object, report our results from the light curve solution and
the spectral analysis, and propose a possible configuration for this system.
\section{Positional information}
\label{sec:positional}
A search with SIMBAD yielded no catalogued object at or near the coordinates of
the eclipsing object, but loading the Incremental Release Extended Source
Catalog of the Two Micron All Sky Survey (2MASS) into ALADIN resulted in a
match. We could clearly identify our object with the point source
\object{2MASS\,J0516288+260738}, and later with a point source in the
USNO-B catalogue (cf. Sect.\,\ref{sec:colour}).
We use the 2MASS catalog entry to give improved coordinates:
\par
RA~=~05$^{\rm h}$16$^{\rm m}$28\fs 81, ~~~~~
$\delta$~=~+26\degr 07\arcmin 38\farcs 8 ~~~~~~~
(J2000).\\
%
\begin{figure}
  \centering
  \includegraphics[width=4.3cm]{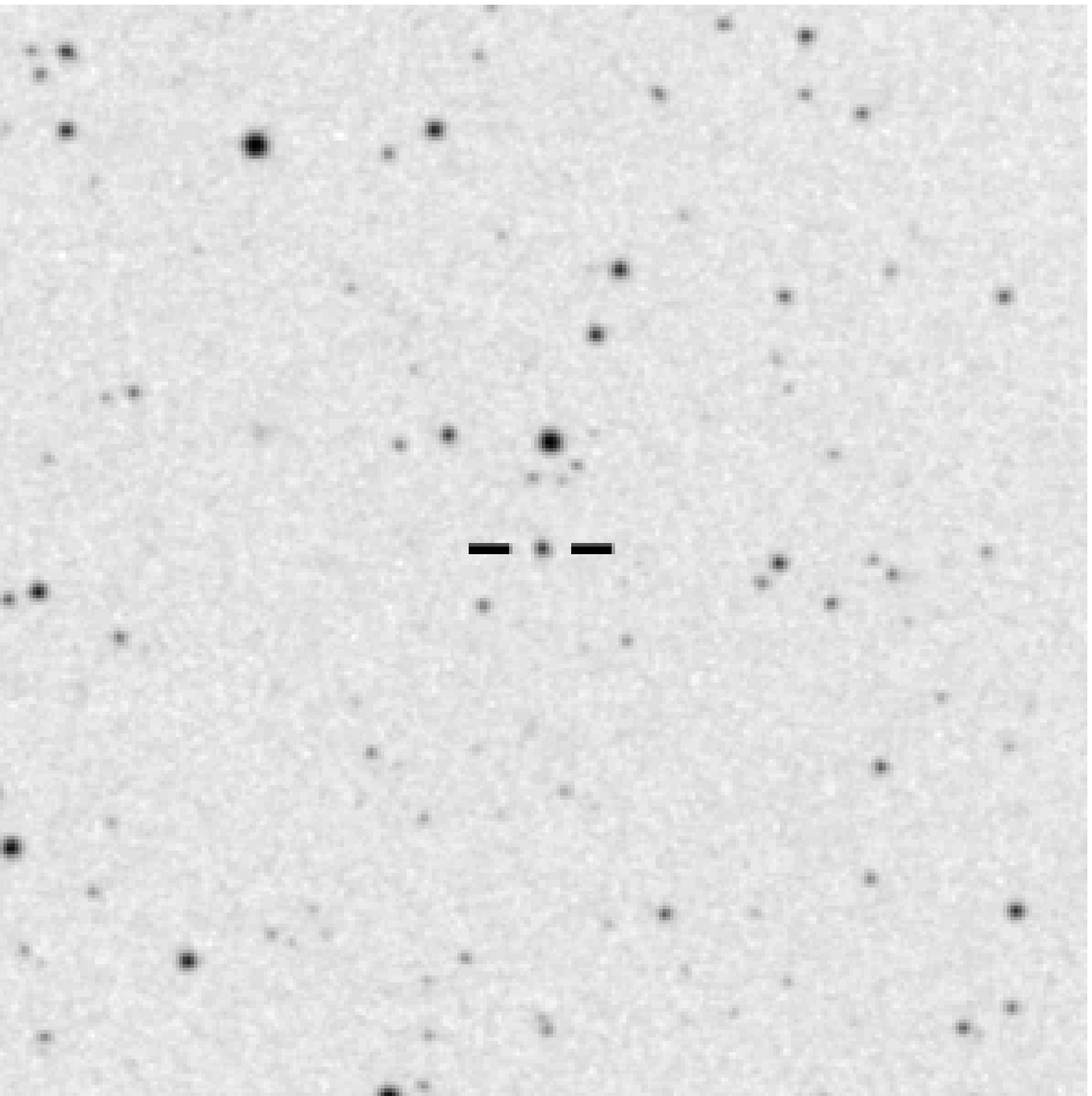}
  \includegraphics[width=4.3cm]{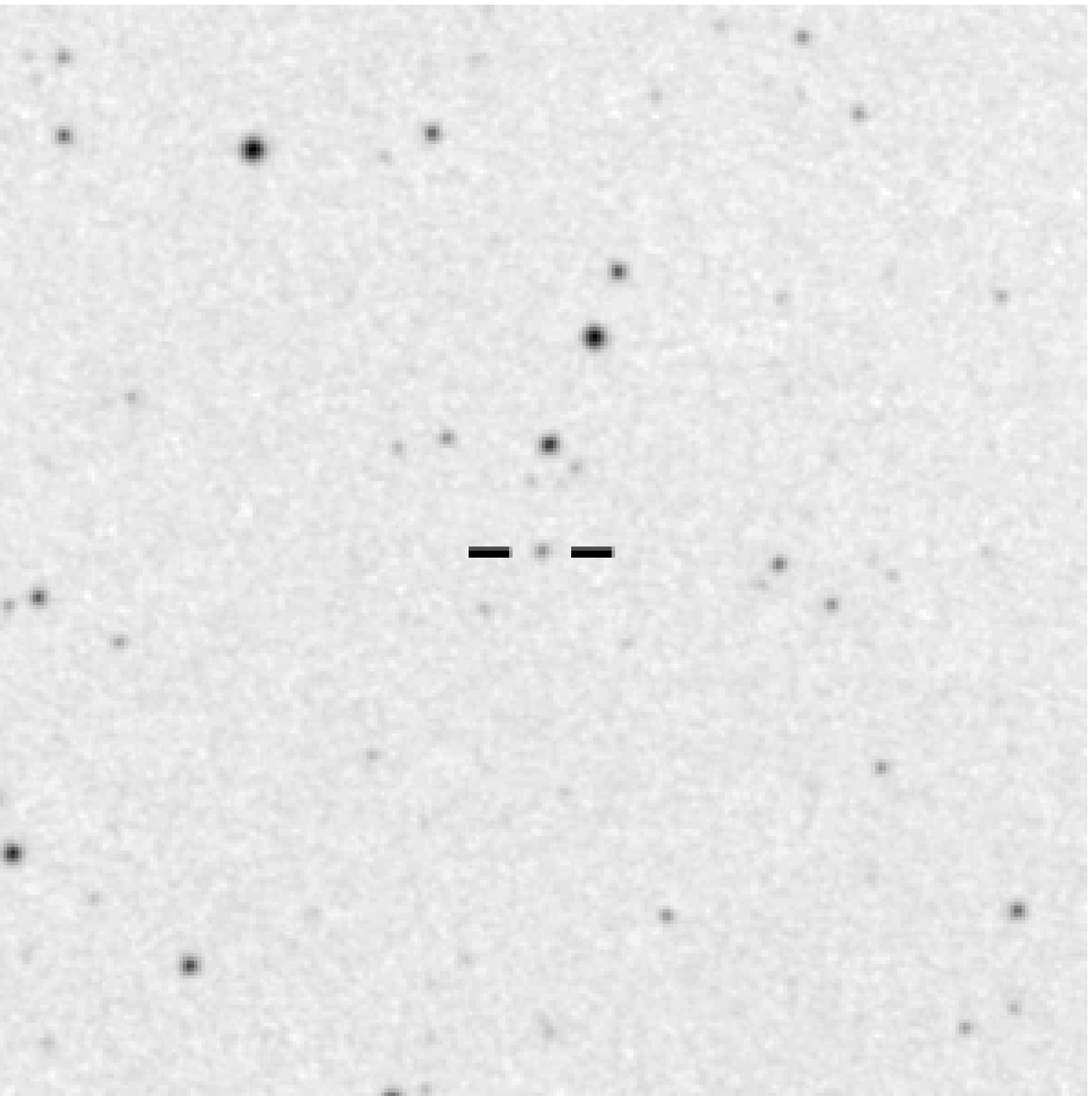}
  \caption{Finding charts for \object{2MASS\,J0516288+260738} (DSS-2
  red: left, DSS-2 blue: right). The side length is 4'x\,4'
  for each image; north is up and east is to the right. 
}
  \label{fig:finding}
\end{figure}
For a clear identification, the object is marked with horizontal bars
in the finding charts given in Figure \ref{fig:finding}.
\section{The light curve}
\label{sec:lightcurve}
\begin{table*}
  \caption[]{Photometric observations.}
  \begin{tabular}{lr@{~}r@{~}rr@{:}lrrrrrl}
    \hline
    \hline
    Site      &  \multicolumn{5}{l}{start time [UT]} &
    length[h]&frames&epoch&minimum [HJD]& O$-$C & \hfill comments\\
    \hline
    SARA      & 2001 & Dec & 06 & 05&01 & 7.1  & 857 & &&&\hfill used for profile\\
    SARA      & 2001 & Dec & 07 & 03&14 & 5.5  & 664 & &&&\hfill used for profile\\
    SARA      & 2001 & Dec & 07 & 10&14 & 1.9  & 233 & &&&\\
    CAHA      & 2001 & Dec & 07 & 20&09 & 8.8  &1032 &0&2452251.5164&$-$0.0009&\hfill used for profile\\
    Piszk\'estet\H o & 2001 & Dec & 07 & 19&23 & 7.3 & 380 &0&2452251.5164&$-$0.0009\\
    BOAO      & 2001 & Dec & 08 & 11&20 & 9.0  &1029 & &&&\\
    Piszk\'estet\H o & 2001 & Dec & 08 & 18&14 & 4.7 & 200 & &&&\hfill used for profile\\
    WISE      & 2001 & Dec & 09 & 01&12 & 2.4  & 690 & &&&\hfill used for profile\\   
    WISE      & 2001 & Dec & 09 & 18&02 & 1.8  & 511 & &&&\\ 
    Piszk\'estet\H o & 2001 & Dec & 09 & 18&26 & 7.5 & 370 & &&&\hfill used for profile\\
    WISE      & 2001 & Dec & 09 & 20&46 & 2.6  & 768 & &&&\hfill used for profile\\ 
    WISE      & 2001 & Dec & 09 & 23&43 & 2.0  & 564 & &&&\hfill used for profile\\ 
    WISE      & 2001 & Dec & 10 & 01&47 & 1.8  & 505 & &&&\hfill used for profile\\ 
    BOAO      & 2001 & Dec & 11 & 16&51 & 1.6  & 181 & &&&\hfill used for profile\\
    BOAO      & 2001 & Dec & 11 & 19&17 & 0.9  & 101 & &&&\hfill used for profile\\
    SAAO      & 2001 & Dec & 11 & 22&22 & 3.0  &1080 &3&2452255.4007&$+$0.0016&\\
    GAO       & 2001 & Dec & 12 & 11&33 & 1.4  & 308 & &&&\\
    SAAO      & 2001 & Dec & 12 & 21&41 & 1.7  & 627 & &&&\\
    GAO       & 2001 & Dec & 13 & 13&47 & 5.3  &1000 & &&&\\
    SAAO      & 2001 & Dec & 13 & 19&57 & 5.4  &1821 & &&&\\
    GAO       & 2001 & Dec & 14 & 12&37 & 6.7  &1301 &5&&&\\
    BOAO      & 2001 & Dec & 14 & 16&35 & 2.5  & 257 & &&&\hfill used for profile\\
    Piszk\'estet\H o & 2001 & Dec & 14 & 18&08 & 6.7 & 328 & &&&\hfill used for profile\\
    SAAO      & 2001 & Dec & 14 & 20&11 & 5.1  &1821 & &&&\\
    SAAO      & 2001 & Dec & 15 & 22&01 & 1.9  & 643 & &&&\\
    GAO       & 2001 & Dec & 16 & 16&25 & 3.2  & 700 & &&&\hfill used for profile\\
    SAAO      & 2001 & Dec & 16 & 19&58 & 5.3  &1901 &7&&&\\
    GAO       & 2001 & Dec & 17 & 10&51 & 0.9  & 203 & &&&\\
    GAO       & 2001 & Dec & 17 & 13&19 & 0.6  & 125 & &&&\\
    GAO       & 2001 & Dec & 17 & 15&50 & 3.2  & 700 & &&&\\
    SAAO      & 2001 & Dec & 17 & 20&02 & 5.1  &1827 & &&&\\
    GAO       & 2001 & Dec & 18 & 10&22 & 8.6  &1750 & &&&\\
    SAAO      & 2001 & Dec & 18 & 20&03 & 4.8  &1710 & &&&\\
    GAO       & 2001 & Dec & 19 & 12&22 & 6.7  & 753 &9&2452263.1615&$-$0.0013&\hfill used for profile\\
    SAAO      & 2001 & Dec & 19 & 20&04 & 4.8  &1738 & &&&\\
    SARA      & 2001 & Dec & 19 & 04&54 & 7.2  & 817 & &&&\hfill used for profile\\
    GAO       & 2001 & Dec & 20 & 10&54 & 6.2  & 650 & &&&\hfill used for profile\\
    SAAO      & 2001 & Dec & 20 & 19&58 & 4.9  &1752 &10&2452264.4599&$+$0.0031&\\
    SARA      & 2001 & Dec & 20 & 02&37 & 9.7  &1107 & &&&\hfill used for profile\\
    SAAO      & 2001 & Dec & 21 & 19&53 & 5.0  &1804 & &&&\\
    GAO       & 2001 & Dec & 22 & 12&50 & 6.0  & 625 & &&&\hfill used for profile\\
    SAAO      & 2001 & Dec & 22 & 21&22 & 3.0  &1074 & &&&\\
    SARA      & 2001 & Dec & 22 & 04&31 & 3.3  & 404 &11&2452265.7487&$-$0.0019&\\
    SARA      & 2001 & Dec & 23 & 05&12 & 0.8  & 100 & &&&\\
    SARA      & 2001 & Dec & 23 & 10&53 & 1.0  & 125 &12&&&\\
    GAO       & 2001 & Dec & 23 & 12&37 & 5.8  & 700 &12&2452267.0450&$+$0.0004&\hfill used for profile\\
    SAAO      & 2001 & Dec & 23 & 19&54 & 4.9  &1663 & &&&\\
    GAO       & 2001 & Dec & 24 & 12&37 & 6.0  & 720 & &&&\hfill used for profile\\
    CAHA II   & 2002 & Oct & 31 & 23&11 & 3.6  &  50 & &&&\hfill used for profile\\
    CAHA II   & 2002 & Nov & 04 & 22&43 & 1.8  &  27 & &&&\hfill used for profile\\
    CAHA II   & 2002 & Nov & 07 & 02&14 & 2.4  &  17 &259&2452586.6506&$+$0.0010&\\
    CAHA II   & 2002 & Nov & 09 & 23&02 & 6.9  &  99 & &&& \hfill ~~~~
                                                     Johnson I filter data\\
    CAHA II   & 2002 & Nov & 11 & 22&59 & 5.2  &  75 &262&2452590.5304&$-$0.0010& \hfill  
                                                     Johnson I filter data\\
    \hline
  \end{tabular}
  \label{tab:observations}
  \caption[]{Key to observatory sites.}
  \begin{tabular}{llll}
    \hline
    \hline
    Site  &  & Telescope & Observers\\
    \hline
    CAHA  & Calar Alto Observatory, Centro Astron\'{o}mico Hispano Alem\'{a}n, Almer\'{\i}a, Spain & 2.2\,m & SD, SLS\\
    CAHA II&Calar Alto Observatory, Centro Astron\'{o}mico Hispano Alem\'{a}n, Almer\'{\i}a, Spain & 1.2\,m & TN, EG\\
    BOAO  & Bohyunsan Optical Astronomy Observatory, Korea                     & 1.8\,m & SLK, BGP\\    
    SAAO  & South African Astronomical Observatory, Sutherland, South Africa & 1.0\,m & GH, TM\\
    SARA  & 
    Kitt Peak National Observatory, Tucson, Arizona, United States of America & 0.9\,m & MW\\
    GAO   & Gunma Astronomical Observatory, Japan                        & 1.5\,m & OH, KK, HT, EK\\
    WISE  & The Florence and George Wise Observatory, Tel-Aviv University, Israel & 1.0\,m & EL, PI, YL\\
    Piszk\'estet\H o & Piszk\'estet\H o, the mountain station of Konkoly Observatory, M\'atra Mountains, Hungary & 1.0\,m & MP, BS, GV, DZ\\
   \hline
  \end{tabular}
  \label{tab:observatories}
\end{table*}
\subsection{Time-resolved photometric data}
All photometric data sets obtained with CCD cameras in the December
2001 \object{KUV\,05134+2605} campaign were compiled. Additional
observations obtained in November 2002 were added later. A list of all
data sets used is given in Table~\ref{tab:observations}, with a
complementary key to the observing sites involved in
Table~\ref{tab:observatories}. All data were bias and flatfield 
corrected according to standard routines. Aperture photometry was
performed on all of these frames using the TRIPP package
\citep{schuh:03d}. Two reference stars that are available on all
frames (shown to be stable during the whole 2001
campaign) were chosen and used consistently for all data sets to
produce relative light curves. The total light curve was then scaled
by a unique factor to produce a light curve with a mean
relative intensity of unity for the white light contributions outside
eclipse. Finally, all times were converted from Julian date (JD) to
heliocentrically corrected Julian date (HJD). 
The result is shown in Fig.\,\ref{fig:lc}. 
\par
The light curve shows a clear periodicity of 1.29 days. All observed
eclipses are similar to each other, have a duration of about 0.10 days and
exhibit a decrease in flux of 15\% (or 0.17\,mag) at the deepest point. There is no
indication of a secondary eclipse in any of the eight 2001 data sets
that partly or fully cover the phase where such an event would be
expected. Furthermore, the three 2002 data sets covering that phase put a
clear upper limit to the depth of any secondary eclipse:
0.49\,\% (or 5.4\,mmag) in white light and 0.70\,\% (or 7.6\,mmag) in the Johnson I filter.
\subsection{Ephemeris}
\label{sec:ephemeris}
Primary minima times were determined by fitting parabolas to the
eclipses. The results for the epochs 0 (two independent data sets), 3,
9, 10, 11, 12 (concatenated from two different non-overlapping data
sets), 259 and 262 are given in Table~\ref{tab:observations}; no fits
could be obtained for epochs 5 and 7 
since only parts of either ingress or egress had been observed there. A
linear regression for the measured minima times then gives the linear
elements and their 1$\sigma$ errors for the primary minima as
\[ \begin{array}{r@{}lcr@{}l}
  {\rm HJD} = 2452251\fd 51&73  & + &    1\fd 293&95 \cdot E. \\
                    \pm &16  &   &    \pm  &25 \\
\end{array} \]
This ephemeris was used to generate a folded profile from the data
taken in 2001.
\par
The folded profile has also been carefully inspected to verify that no
secondary eclipse is apparent in the data. The profile remains at the same
relative flux level outside of the primary eclipse with no significant
indication of ellipsoidal light variations or reflection effects. It
was then used to obtain a light curve solution as discussed in
Sec.\,\ref{sec:lcsolution}.
\begin{figure*}
  \includegraphics[width=16cm]{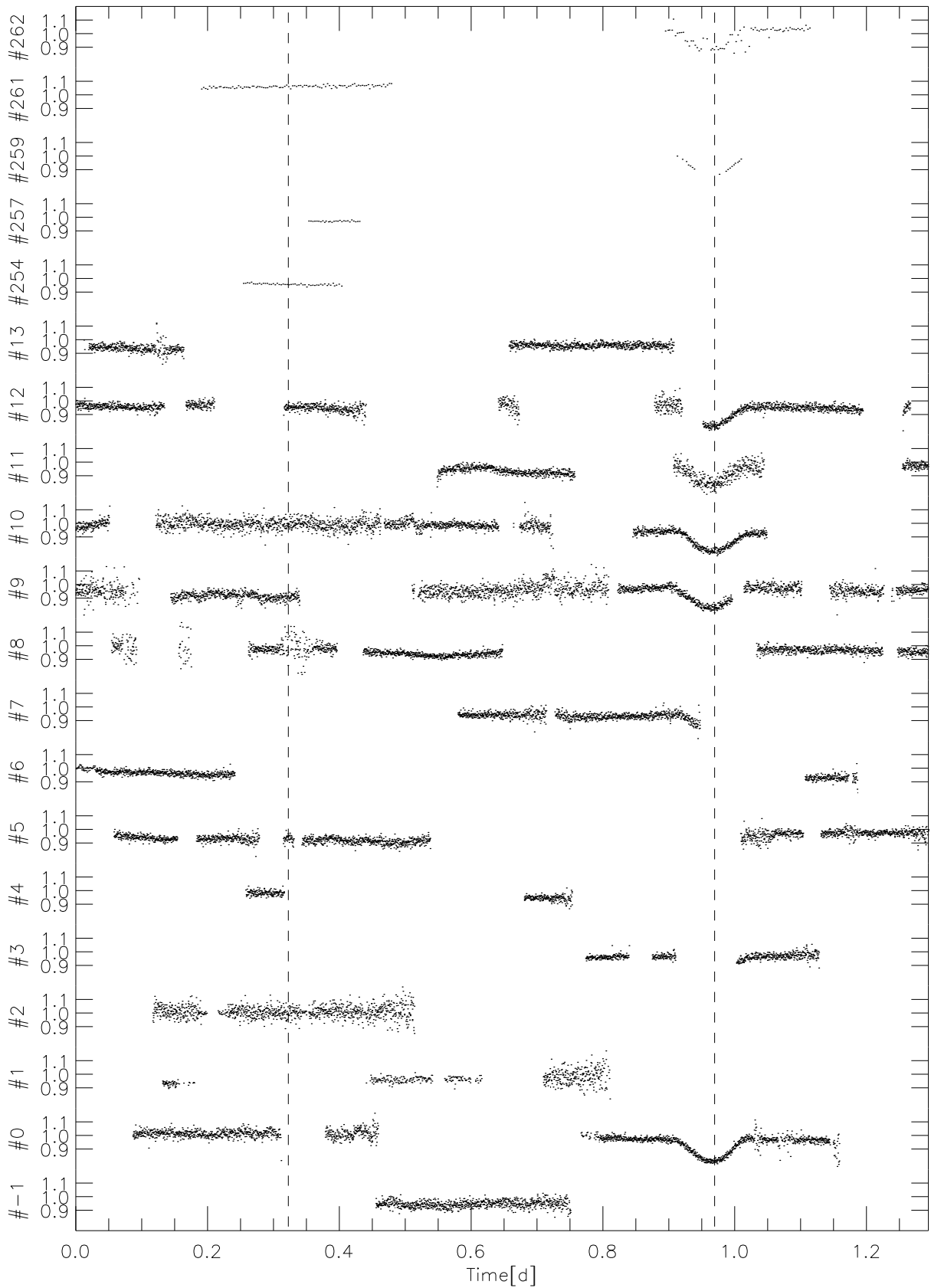}
  \caption{Overview of the photometric observations; the 
    flux is given in fractional intensity units. Time is in days, with
    data binned into units of 30\,s. Time increases from left to right and
    from bottom to top. The epochs (labelled \#n) are displayed
    continuously up to the end of the 2001 campaign, while for the
    2002 observations only those epochs were included in the plot for
    which data points exist. The primary eclipse is displayed
    at multiples of 0.97\,d to place it at a phase of 0.75 in this
    plot, allowing convenient viewing of both phase 0.25 where the
    secondary minima would be located as well as of the primary
    eclipses (both marked by horizontal dashed lines). The scatter in
    the individual light curves contributed by different sites varies
    according to aperture, actual exposure time and weather conditions.}
  \label{fig:lc}
\end{figure*}
\section{Colour and spectral information}
\label{sec:colour+spectrum}
\subsection{Colours}
\label{sec:colour}
To derive a visual magnitude for the analysed object, we have obtained
further photometry of the field. The observations were made at the
South African Astronomical Observatory's 30''~telescope using the UCT
CCD Photometer. A series of 10 images was taken in a Johnson V filter 
on March 10 2003 starting at 18:19:03 UT (exposure time 100\,s).
The measurements were thus made out of eclipse.
We compare our results for \object{KUV\,05134+2605} and
\object{2MASS\,J0516288+260738} to estimate that the
mean V magnitude difference (KUV~$-$~2MASS) is 1.44$^m~\pm$~0.02.
Due to the rapid nature of the variations in the DB variable, it is
not a problem to derive a good mean magnitude for that object. Using
V=16.70 \citep{1990AJ.....99.1907W}) for \object{KUV\,05134+2605}, we
obtain V$_{\rm 2MASS}~=~18.1^m~\pm$~0.1.
\par
Photographic B, R and I magnitudes are published in the 
USNO-B catalogue \citep{2003AJ....125..984M}, and 
infrared J, H and K magnitudes are available from the 2MASS
catalog. They are compiled in Table~\ref{tab:2mass}.
The USNO-B and 2MASS magnitudes $m$ have been converted to
$F_{\lambda}$ using the relation
\begin{table}[h!]
  \caption[]{Johnson V, USNO-B B, R, I and 2MASS J, H, K magnitudes.}
  \label{tab:2mass}
  \begin{tabular}{@{}lc@{}c@{}c@{}c@{}c@{}c@{}c@{}c}
    \hline
    \hline
              & B & V & R & I& J     & H          & K         \\
    \hline
    magnitude &19.47&18.1&16.8&15.84&14.247&13.346&13.115\\
    error     &$\pm$0.3~&~$\pm$0.1~&~$\pm$0.3~&~$\pm$0.3~&~~$\pm$0.040~&~~$\pm$0.039~&~~$\pm$0.039\\
    \hline
    \hline
    $\lambda_{0}~~{\rm [\mu m]}$&0.43&0.55&0.70&0.90&1.25   &1.65        &2.17\\
    $F_{{\nu}_{0}}{\rm [Jy]}$ &4440~~~&~~3810~~~&~~2880~~~&~~2240~~&~~1593~~&~~1089~~&~~713\\
    $F_{{\lambda}_{~}}{\rm [~*~]}$
              &1.17&2.17&3.36 &  3.83  &6.12    &5.51        &2.58\\
    \hline
    \multicolumn{6}{l}{\rule{0cm}{3mm}* 
         ${\rm [erg \,s^{-1} cm^{-2} \AA^{-1} \cdot 10^{-16}]}$}\\
  \end{tabular}
\end{table}
\par
\begin{displaymath}
F_{\lambda} {\rm [erg\,s^{-1} cm^{-2} \AA^{-1}]} =  F_{{\nu}_{0}}{\rm
  [Jy]}\!\cdot\!10^{-0.4 m}\!\cdot\!3\!\cdot\!10^{-13} / \lambda_{0}{\rm [\mu m]}^2.
\end{displaymath}
\par
The photometric zero points $F_{{\nu}_{0}}$ and central wavelengths
$\lambda_{0}$ used for the conversion are
tabulated in Table~\ref{tab:2mass} along with the results for
$F_{\lambda}$.
For the 2MASS filters, these quantities were obtained from
\citet{squires02}, while for the white light we used the values
published for V by \citet{1985AJ.....90..896C}
and \citet{1985AJ.....90..900R} for the Johnson UBVRI+ system.
\par
The Sloan Digital Sky Survey (SDSS) does not cover the field in its EDR (Early Data
Release, \citealt{2002AJ....123..485S}) 
so that no further photometric information is available. Since the object,
according to its infrared colours, is very red, we also checked the VLA
FIRST survey at 20\,cm, which currently does not cover this field either,
and the NRAO/VLA Sky Survey (NVSS) at 1.4 GHz \citep{1998AJ....115.1693C}, 
which covers the field but does not show a radio source in the vicinity.
For completeness, we finally note that neither the ROSAT Bright Source
Catalogue as compiled from the WFC All Sky Survey \citep{1993MNRAS.260...77P} nor the
ROSAT XUV Pointed Phase Source Catalogue as compiled from WFC observations
during pointed phase \citep{1995A&AS..114..465K} list sources at or near the object's position.
\par
\subsection{Optical spectra}
Two medium resolution spectra of \object{2MASS\,J0516288+260738} were obtained in
February 2002 at the Calar Alto 3.5\,m telescope with the double beam
spectrograph TWIN (see Table~\ref{tab:specobs}, first part). Gratings
\#\,5 and \#\,6 were used for the blue and red arm, respectively, with
the dichroic set at 550\,nm. Together with slit widths of 1\farcs 2
and 1\farcs 5 for the first and second exposure, this resulted in spectral
resolutions of 0.94 and 1.04 \AA.
Both spectra turned out later on to have been taken well outside any
eclipse, but only the spectrum taken on February 25 reaches an
exposure level acceptable for further analysis: the signal-to-noise
level per pixel for the first spectrum is only 3, but reaches 8 for the second
one. The frame and corresponding wavelength calibration frame were
bias and flatfield corrected. Then the spectrum was extracted, sky
corrected, subjected to a cosmic ray filtering, corrected for the
illumination function, and finally wavelength calibrated. 
Flux calibration was done by first applying the same procedure to an
exposure of the standard star G191-B2B taken in the same night, then
using tabulated flux values to do the absolute calibration.
The two resulting optical spectra in the wavelength ranges of
3900\,--\,5000\,\AA\ and 6000\,--\,7090\,\AA\ of
\object{2MASS\,J0516288+260738} are displayed in Fig.\,\ref{fig:flux}
(rescaling as described in Sect.\,\ref{sec:interpretation}).
At the given S/N ratio, no lines or features, in
particular no TiO bands, can be discerned.
\begin{table*}[ht!]
\begin{center}
  \caption[]{Spectroscopic observations}
  \label{tab:specobs}
  \begin{tabular}{lllr@{~}r@{~}rr@{:}lrr}
    \hline
    \hline
    object &instrument&$\lambda$ range [\AA]& 
    \multicolumn{5}{l}{start time (UT)} & exptime & HJD \\
    \hline
    \object{2MASS\,J0516288+260738}&TWIN&3900\,--\,5000, 6000\,--\,7090
               &2002 & Feb & 23 & 21&34& 1800\,s & 2452329.411\\
    G191-B2B   &TWIN&3900\,--\,5000, 6000\,--\,7090
               &2002 & Feb & 25 & 18&40&  300\,s & 2452331.281\\
    \object{2MASS\,J0516288+260738}&TWIN&3900\,--\,5000, 6000\,--\,7090 
               &2002 & Feb & 25 & 20&49& 1800\,s & 2452331.380\\
    \hline
    \hline
    \object{2MASS\,J0516288+260738}&OMEGA-Cass&HK: \hfill 14000\,--\,25000
               &2002 & Oct & 27 &23&31&24\,$\times$\,120\,s& 2452575.484\\
    GD\,71     &OMEGA-Cass&HK: \hfill 14000\,--\,25000
               &2002 & Oct & 28 &02&21&10\,$\times$\,120\,s& 2452575.602\\
    \hline
    \hline
    \object{2MASS\,J0516288+260738}&OMEGA-Cass&JH: \hfill 10000\,--\,18000
               &2003 & Feb & 03 &21&49&24\,$\times$\,120\,s& 2452674.412\\
    GD\,71     &OMEGA-Cass&JH: \hfill 10000\,--\,18000
               &2003 & Feb & 03 &23&12&24\,$\times$\,120\,s& 2452674.470\\
    \hline
  \end{tabular}
\end{center}
\end{table*}
\par
\subsection{Infrared spectra}
By the start of the following observing season for
\object{2MASS\,J0516288+260738} in autumn 2002, the compilation and
reduction of the light curve was not only complete enough to allow the
prediction of eclipse times, but also to attempt a first light curve
solution, based on an estimate of the spectral class obtained from the
slope of the optical spectrum. This confirmed the suspicion that the system might
be made up from two low mass stars or a low mass star and a
substellar object, and therefore justified taking infrared
spectra during Director's discretionary time at Calar Alto
Observatory. An H and K band spectrum was observed in October
2002, and a J and H band spectrum in February 2003. 
\par
Each time, a set of 24 spectra was obtained using the OMEGA-Cass instrument
mounted on the 3.5\,m telescope (see Table~\ref{tab:specobs}), well
off both primary and secondary eclipse. Since the 
background is high for infrared observations, the set of 24 spectra
was obtained in such a way that alternating exposures contain the
source on two different locations on the chip. After bias and
flatfield correction of the individual exposures, this allows to
determine a mean background at the (dispersed) location of the source
for both types of images from the respective subset of the
complementary frames.  These two measures of the mean background can
then be used to subtract the appropriate background from all frames of
the two subsets. To do this, the shift for each background row was first
determined by cross-correlating it along the dispersion direction with
the corresponding image row, and the overall run of the shift along
the chip obtained by fitting these row-by-row measurements with a
low-order polynomial. To achieve the best possible subtraction in the
vicinity of the source location on the chip, this fit with sub-pixel
accuracy was then used to shift the background onto the image before
subtracting it. For the wavelength dispersions at the two source
locations, no shifts could be detected during the course of the
exposure series. Therefore, next these bias subtracted, flatfield corrected
and background subtracted frames of each of the two sets were added to
yield two summed images. The spectra were extracted from these two
images using standard procedures for extraction, sky correction,
cosmic filtering, illumination correction, and wavelength
calibration. 
The same procedure was used for the set of 10 and 24 exposures of the
standard star GD\,71. For each observation, the two spectra for the standard star were then
combined and compared to tabulated flux values to obtain the factor
for absolute flux calibration, which was then applied to both the two
object spectra and the two individual standard star spectra. 
\par
Comparing the results for the individual spectra for both stars leads
to the conclusion that the error bars in the resulting combined
spectra must be considered to be of the same magnitude as any
``features'' that one might be tempted to spot.
The same conclusion results from a comparison of the H band parts of
the spectra from the two different observing dates, where most ``features''
are not reproduced. Furthermore, a flux difference by a factor of
about 1.5 between those two independent observations gives an estimate
for the errors in the flux calibration. The rescaled infrared spectra 
for
14000\,--\,25000\,\AA\ and 10000\,--\,18000\,\AA\ are displayed  in
Fig.\,\ref{fig:flux}.
\section{Spectral analysis}
\label{sec:interpretation}
The flux calibrated optical and infrared spectra as
well as the  broad band filter measurements converted to flux
values (diamond symbols) are all displayed together in
Fig.\,\ref{fig:flux}. To obtain a consistent image of the spectral
energy distribution, the magnitude measurements were used to rescale
the spectra where necessary. A unique correction factor was applied to
both optical spectra simultaneously, and a correction factor was
applied to each of the two independent infrared spectra (HK and JH).
\par
The uncertainty in the optical spectrum results since both object and
flux standard star were observed under non-photometric conditions.
The same argument applies to the infrared spectral observations, where
observations from different nights, although both nominally flux
calibrated, result in different flux levels for the overlapping H
band. A consistent adjustment therefore seems justified. Residual
errors may result from the transformation of magnitudes to
F$_\lambda$.
\par
In the following, it will be assumed that the observed spectral energy
distribution consists of light from the primary only; furthermore, for
reasons detailed in Sect.\,\ref{sec:hotV}, the primary will be
presumed to be a late main sequence star. \clearpage 
\begin{figure*}
  \centering
\includegraphics[angle=90,width=14cm]{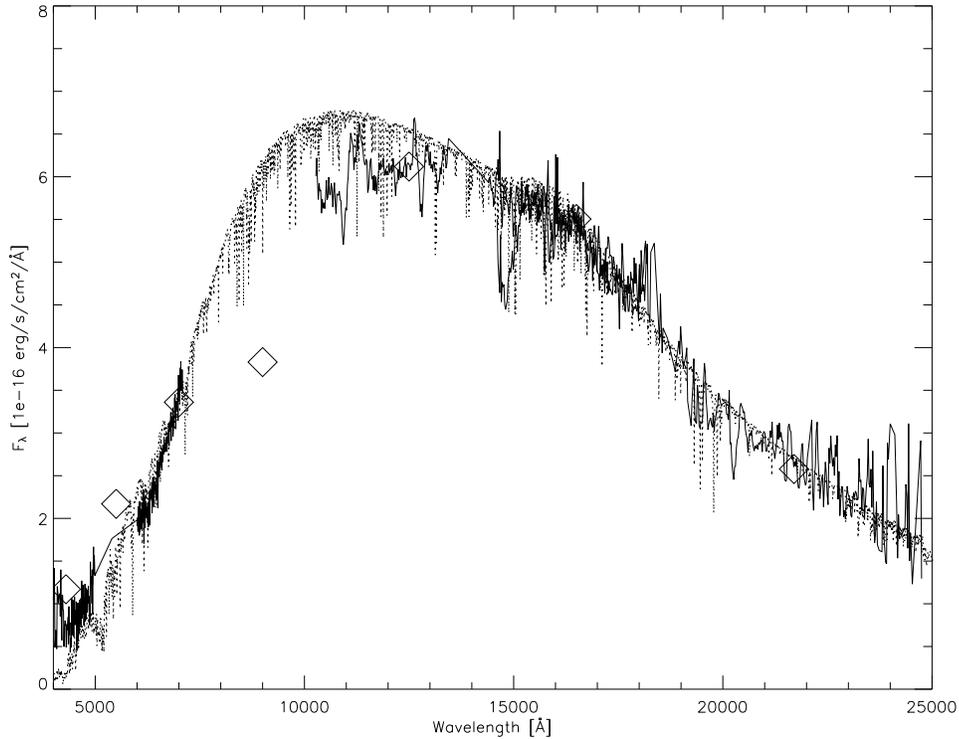}
\caption{Measured flux calibrated optical and IR spectra (solid line),
    USNO-B and 2MASS colours converted to F$_{\lambda}$ (diamond symbols),
    in comparison with the best fit model spectrum (dotted). For details see text.}
  \label{fig:flux}
\end{figure*}
Since
\object{2MASS\,J0516288+260738} is located close to the galactic
plane, the effect of interstellar reddening is not negligible even for
low-luminosity and close-by objects.
\par
For the initial analysis of the observed data we use a grid
of model atmospheres and synthetic spectra that is based on
the models of \cite{2001ApJ...556..357A}. We have extended the 
model grid to effective temperatures of 10\,000\,K for gravities
from $5.5 \le \log{g} \le -0.5$ using spherical symmetry. The 
mixing length was set to twice the pressure scale height, this 
choice of the mixing length was calibrated on early M dwarfs
\cite[]{2002A&A...395...99L}.
\par
Synthetic spectra generated from the models were compared to the observed spectra
using an IDL program. This step was restricted to the infrared spectra.
First, the resolution of the synthetic spectra was
degraded to  that  of  each  observed  spectrum by convolution with
a Gaussian of the appropriate width, and the spectra were normalized to unit
area for scaling.  Next, for each observed spectrum the program calculated a
quality function $q$, similar to a $\chi^2$ value, for the comparison  with all
synthetic spectra  in the grid. The quality function is calculated by first
scaling the model spectrum to the observed fluxes and then by mapping
the synthetic spectrum (reduced to the resolution of the observed data) onto
the grid of observed wavelength points and then calculating
$$
q = \sum_i w_i \left(0.5\frac{f^{\rm model}_i-f^{\rm obs}_i}{f^{\rm model}_i+f^{\rm obs}_i}\right)^2
$$
with $w_i = 0.5(f^{\rm obs}_{i+1}+f^{\rm obs}_{i})(\lambda^{\rm obs}_{i+1}-\lambda^{\rm obs}_{i})$
where $f^{\rm model}$ is the (mapped) flux of the model spectrum,
$f^{\rm obs}$ is the observed flux, and $\lambda^{\rm obs}$ the observed
wavelength. For each model, this procedure was repeated for $0.0 \le E(B-V) \le 2.5$
in steps of 0.1 to independently determine the reddening. For this procedure,
we used the reddening model of \citet[]{1989ApJ...345..245C}.
We then selected the models that resulted in
the 3-10 lowest $q$ values as the most probable  parameter range  for  each
individual star.  The ``best'' value was chosen by  visual inspection,
at this point additionally considering the optical spectra to ensure a
consistent fit. This procedure  allows a
rough estimate of the uncertainty in the  stellar parameters. Note that it does
not eliminate systematic errors in the stellar parameters due to missing,
incorrect or incomplete opacity sources.  The comparison was done for a total
of 377 model atmospheres with solar abundances in the range $2000\,{\rm K}\le
T_{\rm eff} < 5000\,$K and $5.5 \le \log{g}\le 0.0$. Together with the 
search range in extinction this leads to $7539$ combinations that were 
considered in the procedure. With the exception of allowing the 
extinction to vary this is the same procedure that was used in
\cite{2001ApJ...548..908L} and \cite{SandyBinaries}.
\par
The best fitting model has an effective temperature $T_{\rm eff}=4\,200\,$K 
and a reddening of $E(B-V)=0.9$. The low resolution of the data and 
the relative insensitivity of the spectral energy distribution to gravity 
prevent us from determining a value of $\log{g}$, it is clear, however, that
the object is a dwarf rather than a giant. The formal error in effective
temperature is about $\pm 200\,$K and about $\pm 0.2$ for the extinction. 
The low resolution data also prevent detailed metallicity determinations,
and so far only solar metallicities were considered. Overall, the
spectral analysis results suggest a spectral type of about K7~V
($\pm$ 2 subclasses). 
\par
The resulting fit is shown in Fig.\,\ref{fig:flux}. We have applied the reddening to
the synthetic spectrum (dotted line) in order to facilitate the comparison without
modifying the data themselves. All available spectral and colour information is 
included in the figure. The fit is in general acceptable, unfortunately data are
missing in spectral regions where they would be extremely useful to test the 
resulting model parameters. 
\par
A consistency check of our solution can be performed by comparing our
measured reddening with the model of the Galactic interstellar
extinction constructed by \citet{1992A&A...258..104A}. First we
estimate the distance from the spectral type -- absolute magnitude
calibration of \citet{SK82}. From their Table~13 we get an absolute
magnitude of $M_V=8.1$ for spectral type K7\,V. With $E(B-V)=0.9$ as
derived above, the dereddened $V$ magnitude is 15.3 (adopting
$R=3.1$). Thus the resulting distance module is 7.2, corresponding to
a distance of 280\,pc. The reddening predicted from the Arenou et al.
\nocite{1992A&A...258..104A} model and the position of
\object{2MASS\,J0516288+260738} ($l=178.8$, $b=-6.9$) amounts to
$E(B-V)=0.48\pm0.24$. The scatter results mostly from the
patchiness of the interstellar medium in this region. Although this
value is somewhat smaller than our measured reddening both values
agree within the error limits. Note that the model of Galactic
extinction provides an upper limit of $E(B-V)=0.72\pm0.36$ for the
reddening at the position of \object{2MASS\,J0516288+260738}. This
limit results from the fact that stars exceeding a certain distance
are above the absorbing dust layers of the Galaxy. This allows us to
rule out highly reddened early type stars (cf.\ also the independent
discussion of this aspect in Sect.\,\ref{sec:hotV} which leads to the
same result).
\par
\section{Light curve solution}
\label{sec:lcsolution}
From the overall photometric data set, a subset of 23 contributions
was chosen to create the profile used for the light curve
solution. The subsets, marked ``used for profile'' in
Table~\ref{tab:observations}, were selected according to their
length, the coverage in phase they contributed to, the filter they
were taken in (=none), and their
reliability and quality with respect to trends. In contrast to the
data shown in Fig.\,\ref{fig:lc}, each of the selected data sets was
then cleaned from suspicious points and normalised at its maximum. A
folded profile with 200 points, with a phase bin width of
0.005 units, and with phase zero set at the minimum of primary
eclipse, was then obtained from these data, and a few remaining
clearly unreliable points were removed.
\par
This light curve, formed of 187 normal points (in intensity 
units), normalized to unity outside eclipse, was subjected to a
numerical solution by the application of the MORO code
\citep{1995A&A...294..723D}.
The code is based on the \nocite{1971ApJ...166..605W} Wilson-Devinney
(1971) logistical approach, but incorporates a modified Roche model to
account for radiative interaction between the components and uses the
SIMPLEX method as parameter optimization algorithm.
\par
The solution mode was chosen such that no a priori restriction of the
system configuration was imposed (equivalent to the original
\nocite{1971ApJ...166..605W} Wilson-Devinney mode 2). The total number
of light curve parameters for a single passband curve amounts to
17. Since the observed eclipse minimum depth is only moderate
($\approx 16\%$ of maximum light), as no signature of the secondary
except its light blocking effect is evident, and because no colour information
follows from the white light curve, solutions tend to be
underdetermined, especially if the adjustable parameter set is too
large.  Hence it was important to use any available secondary
information from spectroscopy or stellar atmospheres' theory to reduce
the number of free light curve parameters and keep some of them at
fixed values.
\par
No information at all is available for a possible eccentricity of the orbit,
since the position of the unobserved secondary eclipse cannot be determined,
and radial velocity measurements do not exist so far. Therefore
circular orbits ($e = 0$) and synchronously rotating components were
assumed -- as is mostly the case in close binary systems due to their very short 
synchronization time scales. According to the late spectral type, 
bolometric albedos $A_1$ and $A_2$ were fixed at their usually expected
values of 0.5 for convective outer layers, and gravity darkening exponents
$g_1$ and $g_2$ were set to 0.32 as predicted by Lucy's law (1967). 
\nocite{1967ZA.....65...89L}
Linear limb darkening coefficients are poorly known for very late spectral
types. From an extrapolation of the grids of \citet{1985A&AS...60..471W} and
\citet{1995A&AS..110..329D} at their cool ends one obtains approximate
values of $x_1 = 0.5 - 0.6$, which were used in the solutions.
Values of $x_2$ (and $g_2$) are irrelevant due to the absence of any
measurable secondary light.
\par
The primary effective temperature was always fixed at $T_1=3\,000$~K,
typical for spectral type M5~V, since the result of $T_{\rm
  eff}=4\,200$~K from the spectral analysis has only become available
recently, following the February 2003 OMEGA-Cass observations.
This choice is however not critical, 
because the light curve solution only allows to derive the temperature 
ratio $T_1/T_2$, and from a single unfiltered curve no colour information 
can be extracted. The remaining set of adjustable parameters therefore 
comprised inclination $i$, mass ratio $q=M_2/M_1$, secondary temperature 
$T_2$, surface potentials $\Omega_1$ and $\Omega_2$, primary luminosity 
$L_1$, and third light $l_3$. $L_2$ was not independently adjusted, but 
recomputed from $T_2$ and the secondary surface area over the Planck law. 
Trial runs showed that the percentage of third light $l_3$ attributable to 
a possible unresolved field star tended toward zero (except for solution g, 
see Table~\ref{SOLUTION}), so that this parameter was subsequently fixed at $l_3 = 0$ 
in the iterations of all other solutions.
\par
Convergent solutions were achieved after numerous trial runs with a variety 
of start parameter sets (start simplices) and different parameter 
increments as starting points of the automatic iteration process, which 
covered essentially the whole range of physically reasonable parameter 
values. For reasons discussed earlier the numerical process could not be 
expected to yield a single best and unique solution. Instead, 
for a couple of comparably good solutions, 
there was no obvious way to qualify one of these as definitely best representation,
as judged from the final standard deviations of normal points from the synthetic
curves. 
To give an impression of the typical scatter of final parameters we
present a subsample of 8 different solutions with the relatively best
sigma standard deviations in Table~\ref{SOLUTION}. These are sorted in a
sequence of increasing $q$ values. It is obvious that
one can identify two groups of solutions according to the value of the
mass ratio: solutions a-d cluster around $q \sim 0.10 \pm 0.02$,
while cases e-h yielded $q \sim 0.18 \pm 0.01$.
\par
A common feature of all solutions are consistent values of 
inclination ($i \sim 72\degr-75\degr$), temperature ratio ($T_1/T_2 \sim
1.6-1.8$), ratio of radii ($r_1/r_2$ around 0.9), extremely low secondary luminosity
($L_2/L_1 \approx 1-2 \cdot 10^{-3}$), and similar system configuration.
As shown for solution c in
Fig.\,\ref{fig:solutiona_roche}, which can be
considered as representative for the group of solutions with $q \approx
0.1$, the secondary is of about the same size as the primary, and nearly fills
its Roche lobe in a close to semi-detached configuration. The photometric
determination of $T_2$ and hence the temperature ratio must be considered very
uncertain, because of the
missing secondary eclipse and the extreme luminosity ratio.
\par
The overall representation of the observations by the theoretical light 
curve is very good. Figure\,\ref{fig:solutiona}
(top) displays the normal points in comparison with 
the synthetic curve (solid line). Especially the eclipse minimum is matched 
in detail. The standard deviation amounts to only 7.5~mmag, which 
corresponds to the typical scatter of measurements binned to normal points.
As shown in the bottom part of Fig.\,\ref{fig:solutiona},
most observations lie in a $1\sigma$ 
band, and all within a $3\sigma$ belt, with no apparent systematic 
deviations.
Figure\,\ref{fig:solutiona_triple}
gives a 3-dimensional impression of the aspects of the system at 
different phase angles as viewed under an inclination of $74\degr$; the 
configuration corresponds to the parameters of solution c.
\par
\begin{table*}[h!]
\caption{MORO solutions of the light curve of the eclipsing system
         2MASS J0516288+260738}
\label{SOLUTION}
\begin{tabular}{lllll|llll}
\hline\hline\noalign{\smallskip}
Parameter       & a & b & c & d & e & f & g & h     \\[1mm]
\hline\noalign{\smallskip}
$q~(=M_2/M_1)$
                & 0.085 & 0.094 & 0.110 & 0.122 & 0.156 & 0.181 & 0.182 & 0.184          \\
$i$
                & 75\fdg5 & 74\fdg3 & 74\fdg0 & 73\fdg9
                & 73\fdg5 & 72\fdg4 & 72\fdg9 & 72\fdg3                                  \\
$T_1/T_2$
                & 1.67 & 1.72 & 1.74 & 1.73 & 1.77 & 1.63 & 1.83 & 1.88                  \\
$r_1/r_2^{\,a}$
                & 1.11 & 0.89 & 0.85 & 0.87 & 0.85 & 0.67 & 0.73 & 0.64                  \\
$\Omega_1$
                & 5.138 & 5.350 & 5.510 & 5.520 & 5.598 & 6.123 & 6.001 & 6.220          \\
$\Omega_2$
                & 1.961 & 1.943 & 1.997 & 2.055 & 2.182 & 2.197 & 2.236 & 2.198          \\
$L_1^{b}$
                & 0.998 & 0.998 & 0.999 & 0.998 & 0.999 & 0.993 & 0.999 & 0.999          \\
$x_1^{c}$
                & 0.60$^{*}$   & 0.60$^{*}$   & 0.50$^{*}$   & 0.50$^{*}$
                & 0.60$^{*}$   & 0.60$^{*}$   & 0.488        & 0.60$^{*}$                \\
$x_2^{\,c}$
                & 0.50$^{*}$   & 0.50$^{*}$   & 0.50$^{*}$   & 0.50$^{*}$
                & 0.50$^{*}$   & 0.50$^{*}$   & 0.527        & 0.50$^{*}$                \\
$l_3^{d}$
                & 0\,\%$^{*}$  & 0\,\%$^{*}$  & 0\,\%$^{*}$  & 0\,\%$^{*}$
                & 0\,\%$^{*}$  & 0\,\%$^{*}$  & 1.5\,\%      & 0\,\%$^{*}$               \\[1mm]
\hline\noalign{\smallskip}
$1\sigma$ deviation
                &0.00750&0.00750&0.00749&0.00749&0.00749&0.00755&0.00750&0.00749         \\[1mm]
\hline\noalign{\smallskip}
\multicolumn{9}{l}{$^{a}${Ratio of mean Roche radii;}}\\
\multicolumn{9}{p{13cm}}{$^{b}${Relative luminosity $L_1/(L_1 + L_2)$; $L_2$ is not
    independently adjusted, but recomputed from $r_2$ and $T_2$};}\\
\multicolumn{9}{p{13cm}}{$^{c}${Linear limb darkening coefficient; theoretical
                                 value for $V$ band taken from \citet{1995A&AS..110..329D};}}\\
\multicolumn{9}{l}{$^{d}${Fraction of third light at maximum;}}\\
\multicolumn{9}{l}{$^{*}${fixed.}}\\
\end{tabular}
\end{table*}
\par

\begin{figure}
  \includegraphics[width=9cm]{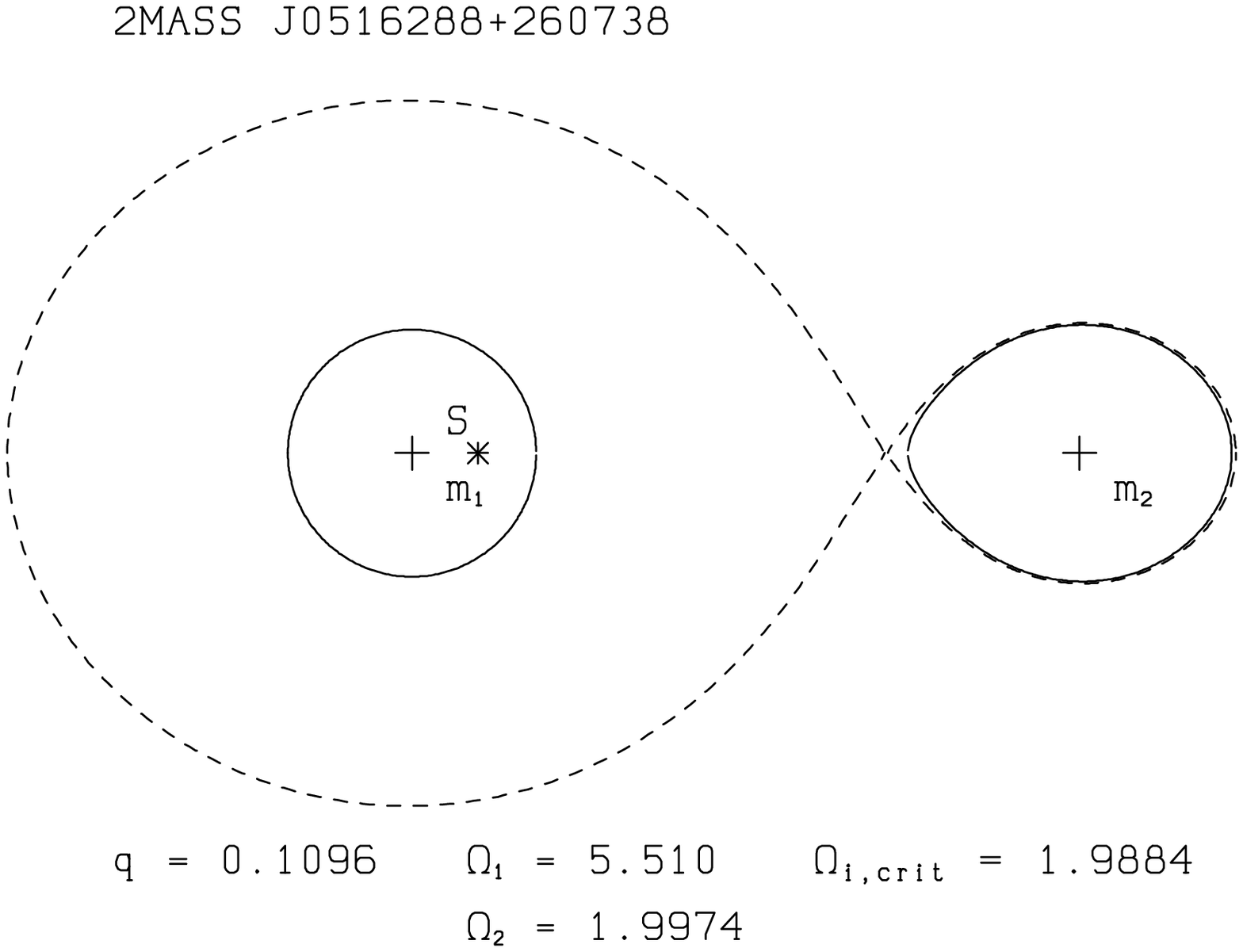}
  \caption{Meridional intersection of surface and inner critical Roche
    equipotentials corresponding to the nearly semi-detached system
    configuration of solution c (see Table~\ref{SOLUTION}); the substellar
    secondary component is close to contact with its Roche lobe.}
  \label{fig:solutiona_roche}
\end{figure}
\begin{figure}
  \includegraphics[width=9cm]{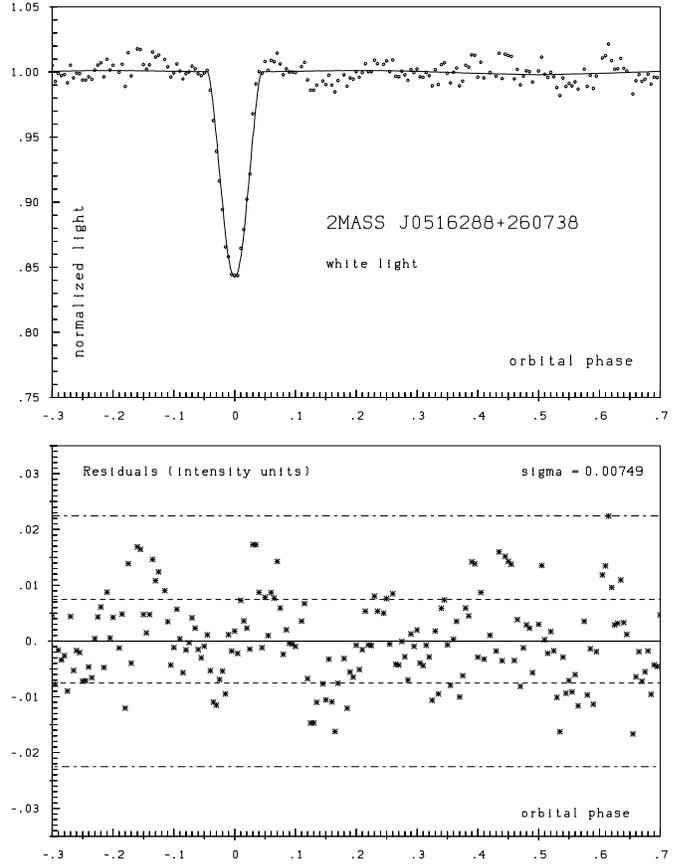}
  \caption{Top part shows the observed light curve in white light
    (dots are normal points formed by binning individual observations to phase
    intervals of width 0.005) together with the theoretical curve
    (solid line) corresponding to solution c of Table~\ref{SOLUTION}; maximum light
    (intensity) was normalized to unity, and phases were computed
    according to the ephemeris of Sect.\,\ref{sec:ephemeris}; bottom part shows
    residuals of observations (in intensity units) with $1\sigma$ and
    $3\sigma$ belts.}
  \label{fig:solutiona}
\end{figure}
\begin{figure}
  \begin{center}
    \includegraphics[width=6cm]{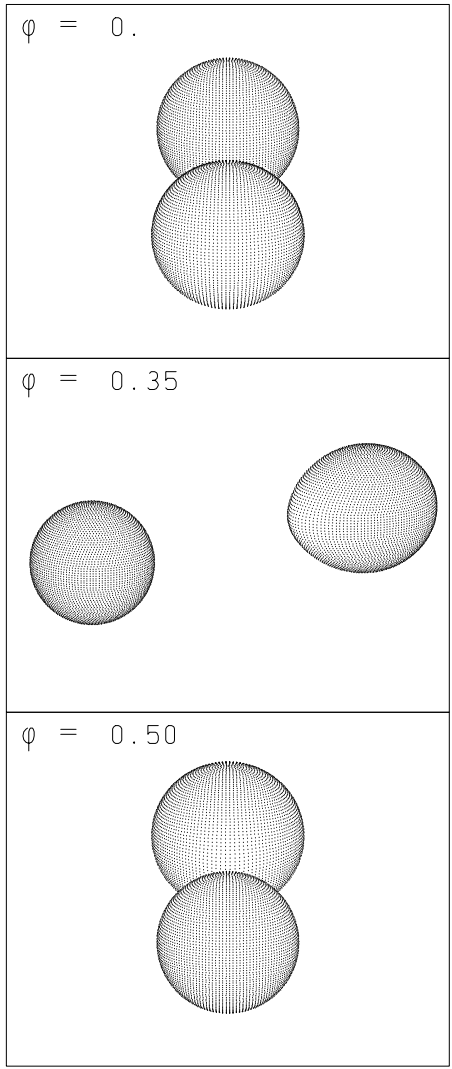}
    \caption{Aspects of the system at three different phases; viewing
      angle is $74\degr$, and system parameters correspond to solution c
      of Table~\ref{SOLUTION}.} 
    \label{fig:solutiona_triple}
  \end{center}
\end{figure}
\begin{figure}
  \includegraphics[width=9cm]{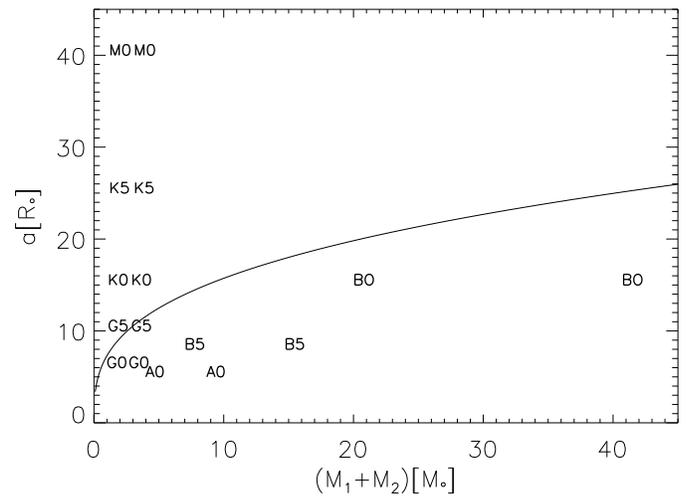}
  \caption{Orbital separation \textbf{a} as a function of the mass sum
    \textbf{$M_1+M_2$} of the system at
    the 1.29\,d period. Overplotted are radii
    for type III giants; explanation see text
    (Sect.\,\ref{sec:hotIII}).}
  \label{fig:exclude}
\end{figure}
\section{Discussion of alternate configurations}
The system is located only 6.9$^{\circ}$ above the galactic plane (in
an outward direction). This implies that the reddening through
interstellar extinction is potentially very high. Although from spectral
observations in conjunction with detailed Galactic extinction models many
stellar spectral and luminosity types other than late main sequence
stars can be excluded, it is also instructive to make use of the independent
information from the light curve solution alone. Through the
geometry of the system, and fundamental stellar parameters that cannot
be substantially altered even when a star resides 
in a close binary system, most of the following alternative
combinations can be excluded. This in turn justifies the restriction of the 
discussion in Sect.\,\ref{sec:interpretation} to a late dwarf system.
\subsection{Reddened giant stars}
\label{sec:hotIII}
Up to spectral types G5 or earlier, luminosity class III red giants
have radii larger than the orbital separation dictated by the measured
orbital period and a total mass sum of the system of up to twice their
own mass. This is illustrated in Fig.\,\ref{fig:exclude}: The solid
line represents the orbital separation of the system components as a
function of the total system mass for the given orbital period of 1\fd
29. It is therefore a strict upper limit to the radius of any single
component of the system. The type III giant star spectral types are
printed at the position of their radii, once over their
corresponding stellar mass and once at twice that value. These
overplotted radii for type III giants were taken from the mass-radius
relation by \citealt{2000asqu.book.....C}.  Using the stellar mass and
the double of it means that in between those two values all possible
mass combinations are covered, since the roles of primary and
secondary would simply become reversed if an even greater fraction of
the total mass were attributed to the presumed secondary. The first
case represents the limit in which the mass of the companion is
negligible, so that the total mass is solely made up of the giant's
contribution, while in the second case the mass of the giant amounts
to half of the total mass in the system.
\par
In a na\"{\i}ve consideration, the early type giants could fit
within the orbital separation, even if it is clear that most of the
time they would reach well over half of the total distance. But
although these earlier types could just about fit into the system, it
can easily be shown that their deformation within the Roche potential
would in all cases result in large ellipsoidal light variations, which
are not observed in the actual light curve. In addition, the width of the
observed eclipse minimum would cover a much broader phase range. For
these estimates, the binary eclipse simulation program
\texttt{nightfall} (R. Wichmann, Landessternware Heidelberg, Germany)
which calculates synthetic light curves taking into account the
distortion of the stars in Roche geometry was used.
\par
Since these considerations equally apply to luminosity classes II and
I, and even in a much stricter form there, the luminosity class for
the more luminous object in the system must be V or higher.
\subsection{Reddened earlier main sequence stars}
\label{sec:hotV}
Giant stars do not fit within the prescribed orbit; but what about
bright early main sequence stars that appear reddened by strong
interstellar absorption? Very early main sequence stars have masses
and especially radii similar to those of type III giants, so that,
as above, geometry arguments can be brought forward to rule out a
combination of two very early-type components. This is important, as
results from stellar structure will be utilized to find
physically meaningful pairs in what follows.
\par
The light curve solutions constrain the mass ratio, the radius
ratio and the ratio of the effective temperatures almost
regardless of the absolute value of T$_1$.
Using tabulated values for the masses, radii and effective
temperatures of stellar and substellar objects, the possible
components making up the system can be constrained by requiring that
both of them have parameters reasonably close to those of isolated
main sequence stars or substellar objects. The stellar parameters used
in the following were taken from \citet{2000asqu.book.....C}, those
for substellar objects (for ages ranging from 1\,M\,yrs to 10\,G\,yrs) from
\citet{2000ApJ...542..464C} and \citet{2002A&A...382..563B}.
\par
\begin{table*}[h!]
\caption{Interpretation of Figs.\,\ref{fig:exclude_2} and
  \ref{fig:exclude_3}; for explanations see Sects.\,\ref{sec:hotV} and
  \ref{sec:discussion}.}
\label{tab:discussion}
\begin{tabular}{lll}
\hline\hline\noalign{\smallskip}
possible primary spectral types for solution&group a-d&group e-h\\
\hline\noalign{\smallskip}
from mass-temperature relation (Fig.\,\ref{fig:exclude_2})& upper limit G5 & region G0 - K0, or upper limit M2\\
from mass-radius relation (Fig.\,\ref{fig:exclude_3})& upper limit G5 & upper limit M0\\
\hline\noalign{\smallskip}
combined constraints                & upper limit G5 & upper limit M2\\
consistency with spectroscopy (K7$\pm2$)& fully consistent &marginally consistent at most\\
\hline\hline\noalign{\smallskip}
resulting secondary mass [M$_{\sun}$]   & upper limit 0.11 & upper limit 0.076\\
resulting secondary mass [M$_{\sun}$] for K7 & 0.062$\pm$0.01& 0.105$\pm$0.01\\
\hline\noalign{\smallskip}
\end{tabular}
\end{table*}
The ratios of effective temperatures and masses of the binary
components for the light curve solutions c and g from
Table~\ref{SOLUTION}, which can be considered representative for the two
bulges of solutions clustering around $q$ = 0.10 and 0.18,
are used to find the corresponding effective
temperatures and masses of the secondary as a function of primary mass.
For all main sequence stars, their effective temperatures and those
required by the two representative solutions for the secondary are plotted in
Fig.\,\ref{fig:exclude_2}. The zero-age main sequence objects and the
youngest substellar objects (1\,M\,yrs old) are marked by plus signs
and are connected by a thick solid line; higher age substellar
models are also represented by plus signs, which remain, however, isolated for
clarity. For both solutions a row of squares connected by a solid line 
is shown. The squares correspond to the locations of the secondary in the 
$T_{\rm eff}-M$ diagram, which follow from the temperature and mass of a 
main sequence primary using the temperature and mass ratios of the 
respective photometric solutions c and g. For each square
  plotted, the corresponding error estimates from the typical
  dispersion within each of the two groups are indicated by small dots which
  represent the end points of the associated error bars (not drawn as
  full lines to preserve more clarity in this complex representation).
For the primary a variety of spectral 
types between O5~V and M8~V were considered to cover the full zero-age main 
sequence (plus signs).
When inspecting this figure and the following graph, note that the
plot scale is logarithmic so that 
offsets between curves can be much larger in regions of the plot that
correspond to the upper main sequence than they might intuitively seem.
\par
The curve for the secondary corresponding to solution c
only approaches and
intersects the main sequence at its lower
end and therefore excludes highly reddened hotter main sequence stars
as a possible primary, since the corresponding secondaries cannot
exist. The other line corresponding to solution g starts off close to the main
sequence and comes back
to it earlier than the other one.  As stated above, the
combination
of two upper main sequence stars as a possible solution can be ruled
out, because such extended stars could only reside within the given
orbit if an appreciable distortion of the primary is allowed for,
which would inevitably result in an easily observable ellipsoidal
light variation. Apart from this special case for g on the upper
  main sequence, solutions (discretised in, on average,
  5-subclass steps!) were elected possible whenever the
  error range for such a discrete secondary location intersected the
  stellar or substellar regime. Errors in parameter ratios are
  regularly smaller than the discretisation in spectral classes used,
  so the limits given can be regarded to be accurate to within roughly
  two subclasses.
\par
On the lower main sequence, the earliest possible
spectral types for the primary in the two cases are as listed in
the first line of Table~\ref{tab:discussion}. In case g, K5 and M0
primaries must be excluded.
When the mass-radius relation is taken into account in
addition to the just invoked mass-temperature correlation, these upper limits
can be even further constrained, as will be shown next. 
\par
The ratios of radius and mass for the binary components (also taken
from Table~\ref{SOLUTION}) together with tabulated mass-radius
relations from the same sources as above can be subjected to the same
procedure. The designations in Fig.\,\ref{fig:exclude_3} are analogous
to those in Fig.\,\ref{fig:exclude_2}. The results are also compiled
in Table~\ref{tab:discussion}. The primary can be constrained
  to be of spectral type G5 or later for solution c, or of spectral
  type M0 for solution g. The second line in
  Table~\ref{tab:discussion} lists these limits without any additional
  age constraints that migth be present (see this discussion later).
\par
\par
Combining the constraints from both Fig.\,\ref{fig:exclude_2} and
Fig.\,\ref{fig:exclude_3} yields an overall upper limit for each
solution as listed in Table~\ref{tab:discussion}, line 3. Solution c
allows for a primary no earlier than G5, while solution g
restricts possible primaries to spectral types no earlier than M2.
Spectroscopic results strongly favour the group a-d solution,
  since the overall constraint of G5 for the primary spectral class is
  entirely consistent with the conclusion in
  Sect.\,\ref{sec:interpretation}. For an upper limit of M2, on the
  other hand, it would be hard to claim consistency with the
  spectroscopy results. Table~\ref{tab:discussion} nevertheless
  explores the mass range for the secondary in different scenarios
  (entries in lines 5 and 6).
\par
An additional constraint not taken into account so far is the age of
the system, which for the more likely solutions a-d is restricted to
below 0.01\,Gyr by the mass-radius relation. A young system is also
allowed for by the mass-temperature relation. However, this
corresponds to a lifetime of the system in which the K star will not
have had enough time to attain the zero-age main sequence yet, and
will hence not necessarily have the ZAMS parameters assumed to deduce
these constraints in the first place. This might well limit the
overall usefulness of this discussion, and is a point that will have
to be re-addressed once better data has become available for this
object.
\subsection{Nearly identical components}
A serious objection to the interpretation presented so far emerges if
the orbital period is really twice as long as assumed (see a similar
initial ambiguity for \object{CM\,Dra} where this indeed turned out to
be the correct interpretation in
\citealt{1967ApJ...148..911E}). The eclipse ephemeris given in
  Sect.\,\ref{sec:ephemeris} would then not correspond to the orbital
  ephemeris of the system any more, as assumed throughout the light
  curve analysis in Sect.\,\ref{sec:lcsolution}. Hence, results
  obtained there are not applicable to the current discussion, where identical 
components could then produce undistinguishable primary and secondary
eclipses. In many binary systems, the mass ratio is close to one, so
this is not an altogether implausible, but from a statistical point of
view highly unlikely configuration. This possible
complication can currently not be resolved, since the scenario could
only be conclusively ruled out, or corrobated, with radial velocity
measurements. 
\subsection{Two old white dwarfs}
A further scenario that requires consideration is a system consisting
of two old and therefore very red white dwarfs (the effect of
interstellar reddening cannot contribute significantly here due to the
low intrinsic luminosity of these objects).
It is however not very probable that the mass ratio 
in a double degenerate system is as low as q $\approx$ 0.1. For the
given period, the duration of eclipse would be of order $10^{-3}$
phase units, completely incompatible with observations.
\section{Conclusion}
\label{sec:discussion}
Despite the remaining uncertainties, from the data presented in
this paper it is plausible that the newly discovered eclipsing binary
system \object{2MASS\,J0516288+260738} consists of a late K-type
(pre-)main sequence star as a primary and a substellar object as a
secondary. For the spectral class upper limits derived in
Sect.\,\ref{sec:hotV}, Table~\ref{tab:discussion} also lists the
secondary masses according to the mass ratios given in
Table~\ref{SOLUTION} for each solution.
All of these mass
values, which are close to or below the
substellar limit  of 0.075\,M$_{\sun}$ required for stable
hydrogen burning, represent upper limits.
Taking into account the additional information available from spectral
analysis which favours a spectral type around K7 results in a value of
$\approx$0.06\,M$_{\sun}$ for the secondary's mass. In this case, only
solutions a-d are considered as likely since a spectral type of K7 would not
be consistent with solutions e-h.
A substellar nature of the companion is therefore quite likely: The
unusually low mass ratios in all solutions make the secondary a
good Brown dwarf candidate.
\par
This interpretation should now be checked by trying to confirm the
spectral classification via the detection of spectral lines in new
high resolution, high signal-to-noise optical and/or infrared
spectra. These lines could then also be used to obtain radial velocity
measurements for the system which should eventually provide absolute
masses.
\par
Together with the extensive light curve available, the system has the
potential to provide a new high-quality point for the mass-radius
relation of the lower main sequence (or for pre-main sequence
evolutionary tracks), and the first one obtained from
eclipse measurements for a sub-stellar object.
\clearpage 
\begin{figure*}
  \includegraphics[width=13.cm]{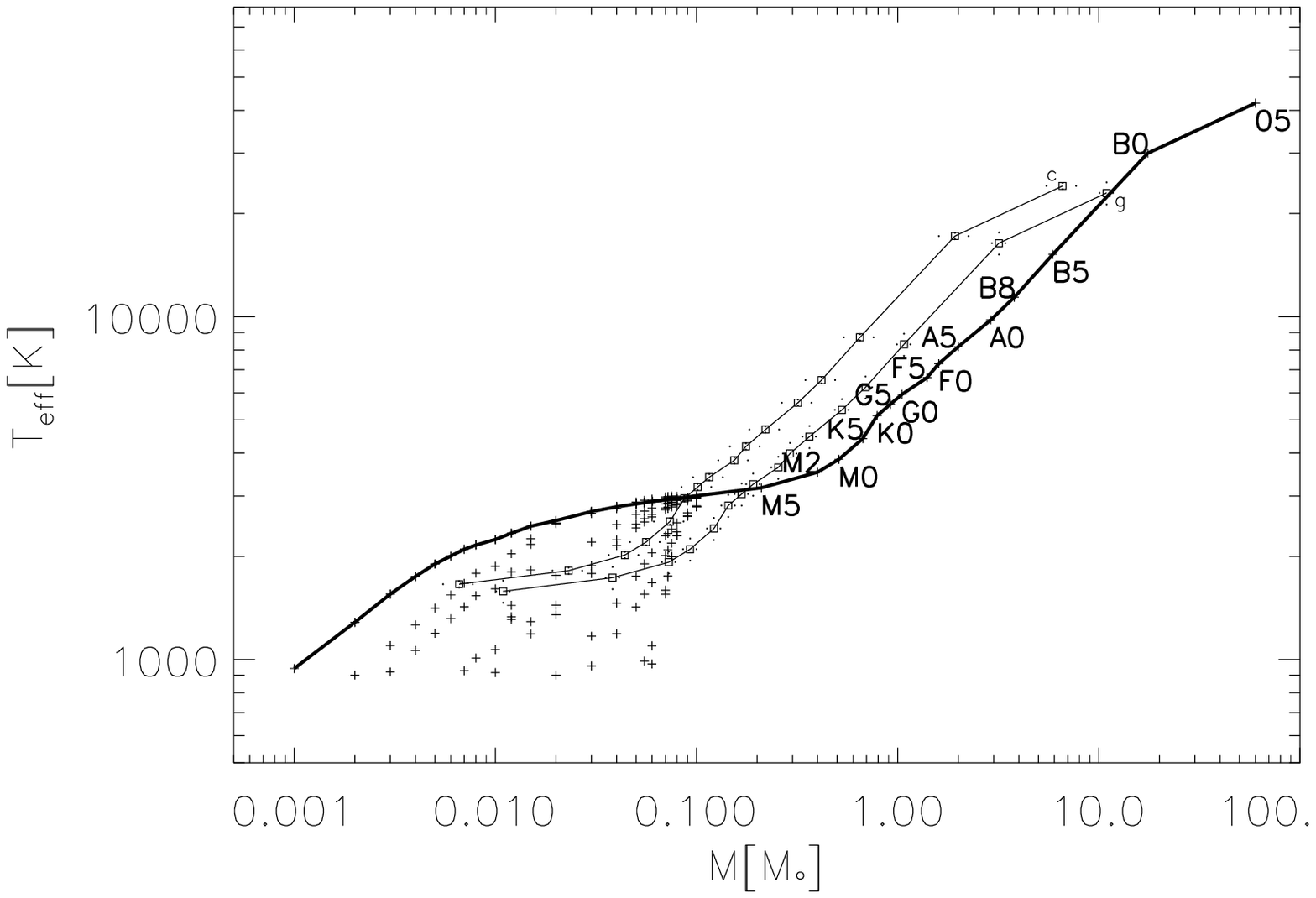}
  \caption{Effective temperature as a function of the stellar mass
    for zero-age main sequence stars and substellar objects
    (plus signs) and their secondaries according to Table~\ref{SOLUTION}
    (squares); further explanations see text (Sect.\,\ref{sec:hotV}).\vspace{2cm}} 
  \label{fig:exclude_2}
\end{figure*}
\begin{figure*}
  \includegraphics[width=13.cm]{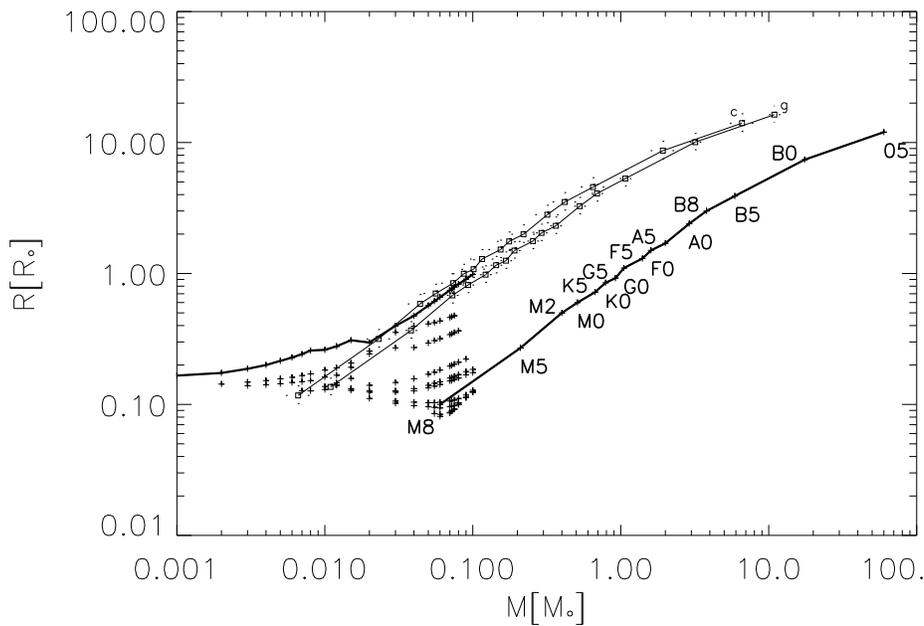}
  \caption{Mass-radius relation
    for zero-age main sequence stars and substellar objects (plus signs)
    and their secondaries according to Table~\ref{SOLUTION} (squares);
    further explanations see text (Sect.\,\ref{sec:hotV}).}
  \label{fig:exclude_3}
\end{figure*}
\par
\clearpage 
\begin{acknowledgements}
  The authors would like to thank K.~Werner and H.~Mauder
  for helpful discussions and friendly support, and P.A.~Woudt for his
  assistence in obtaining a V magnitude for \object{2MASS\,J0516288+260738}.
  We also would like to thank  R.~Gredel for allocating Director's
  discretionary time and U.~Thiele for carrying out the OMEGA-Cass
  observation at Calar Alto observatory in service mode. We
  acknowledge the use of the \texttt{nightfall} program for
  light-curve synthesis of eclipsing binaries
  (http://www.lsw.uni-heidelberg.de/$\sim$rwichman/Nightfall.html), 
  written by R.~Wichmann. Part of this work was supported by the German
  \emph{Deut\-sche For\-schungs\-ge\-mein\-schaft} under project
  grants DR\,281/13-1 and DR\,281/13-2, as well as under travel grants
  DR\,281/16-1, DR\,281/18-1, and NA\,365/6-1. The Wise Observatory
  contribution to this work is supported by the Israel Science Foundation.
  This research has made use of the USNOFS Image and Catalogue Archive
  operated by the United States Naval Observatory, Flagstaff Station
  (http://www.nofs.navy.mil/data/fchpix/).
\end{acknowledgements}

\end{document}